\documentclass{PoS}

\title{High Time Resolution Astrophysics in the Extremely Large Telescope Era : White Paper\thanks{White paper funded as part of  Opticon's HTRA network. Opticon is funded by the EU FP7 programme}}

\ShortTitle{HTRA-WP}

\author{A. Shearer \\ Centre for Astronomy National University of Ireland, Galway, Ireland \\ E-mail: andy.shearer@nuigalway.ie}
\author{G. Kanbach \\ Max-Planck-Institut fuer extraterrestrische Physik, Garching, Germany}
\author{A. S{\l}owikowska\\ Institute of Astronomy Wieza Braniborska  Lubuska Zielona Gora Poland} 
\author{C. Barbieri\\Dipartimento di Astronomia, Universita Degli Studi di Padova, Padova, Italy} 
\author{T. Marsh\\Department of Physics and Astronomy, University of Warwick, UK}
\author{ V. Dhillon\\Department of Physics and Astronomy, University of Sheffield, UK} 
\author{R. Mignani \\Mullard Space Science Centre, University College London, UK} 
\author{D. Dravins \\ Lund Observatory, Lund University, Sweden}
\author{C. Gouiff\'es \\ Service d'Astrophysique, C.E Saclay, France}
\author{C. MacKay\\Institute of Astronomy, University of Cambridge, UK} 
\author{G Bonanno\\ INAF-Astronomical Observatory, Palermo, Italy} 
\author{S. Collins\\ Centre for Astronomy National University of Ireland, Galway, Ireland}

\abstract{High Time Resolution Astrophysics (HTRA) concerns itself with observations on short scales normally defined as being lower than the conventional read-out time of a CCD.  As such it is concerned with condensed objects such as neutron stars, black holes and white dwarfs, surfaces with extreme magnetic reconnection phenomena, as well as with planetary scale objects through transits and occultations. 
HTRA is the only way to make a major step forward in our understanding of several important astrophysical and physical processes; 
these include the extreme gravity conditions around neutron stars and stable orbits around stellar mass black holes. Transits, involving fast timing, can give vital information on the size of, and satellites around exoplanets. In the realm of fundamental physics very interesting applications lie in the regime of ultra-high time resolution, where quantum-physical phenomena, currently studied in laboratory physics, may be explored.
From the short descriptions given below it can be seen that HTRA science covers the full gamut of observational optical/IR astronomy from asteroids to $\gamma$-rays bursts, contributing to four out six of AstroNet's fundamental challenges described in their Science Vision for European Astronomy. 

Giving the European-Extremely Large Telescope (E-ELT) an HTRA capability  is therefore of paramount importance. 
We suggest that there are three possibilities for HTRA and E-ELT. These are, firstly giving the E-ELT first light engineering camera an HTRA science capability. Secondly, to include a small HTRA instrument within another instrument - Micado is a possibility here. Finally, to have separate fibre feeds from, say Optimos, to a dedicated HTRA instrument. In this case a small number of fibres ($<$10) could be positioned and would provide a flexible and low cost means to have an HTRA capability. By the time of E-ELT first light, there should be a number of significant developments in fast detectors, in particular in the infra-red (IR) region - already today small photon-counting IR arrays have been developed that should be mainstream technology in the next years. }

\FullConference{High Time Resolution Astrophysics IV - The Era of Extremely Large Telescopes - HTRA-IV,\\
		May 5-7, 2010\\
		Agios Nikolaos, Crete, Greece}

\begin{document}

\subsection*{Introduction}

The next frontier for observational astrophysics will be the temporal domain below 1 second, where the physics and astrophysics of compact objects becomes important. Traditionally, optical astronomy has been concerned with timescales measured from minutes and hours to years. Sub-minute and particularly sub-second timing has been largely unexplored primarily due to instrumental/detector limitations. However, it is in this regime that there are a number of radically different astronomical phenomena - ranging from transits of extra-solar planets to  $\gamma$-ray burst counterparts at cosmological distances. Expansion of parameter space for astrophysical observations will not only increase the amount of information about well-known (but not necessarily well-understood) phenomena, but will also lead to new discoveries. The expansion of astronomy into new frequency ranges, going far beyond the optical band, has dominated astrophysics in the last decades of the past century. Radio, X and $\gamma$-ray astronomy already routinely observe at timescales down to microseconds  and this has lead to a number of fundamental discoveries. In the ELT era the lack of a full HTRA capability at optical and near-IR wavelengths will restrict a significant number of multi-wavelength studies.

In the time domain below one second observational astrophysics is still rather limited, particularly at optical and near infra-red (NIR) wavebands.  Optical variability longer than, typically, minutes can be measured with standard equipment. To explore phenomena on sub-minute down to sub-$\mu$second time scales, however, requires specialised equipment and high flux levels (i.e. large telescopes). It is now technologically feasible to fully develop access to this astrophysically significant time domain as a result of the parallel advances in detector design and the availability of large telescopes. As a result there are very interesting phenomena being discovered in the optical and at other frequency ranges, combined with theoretical predictions on the physics of very rapid phenomena.

European astronomy is in an excellent position to spearhead this exploration on the basis of existing groups - both theoretical and observational. The proposed HTRA work package will train new researchers in the techniques of this discipline as well as maturing existing collaborations and work programmes. Our scientific objective is to develop the science of HTRA, and specifically:

\begin{enumerate}
\item How will HTRA observations help existing studies and understanding of astronomical objects? 
\item How can we best exploit the current generation of astronomical instruments? 
\item What about future developments - in ten years the next generation of  telescopes will be coming on stream - what HTRA can be done then? 
\item As the E-ELT's optimal wavelength will be around 2.2 $\mu$m what are the detector implications for HTRA in the region 1-2 $\mu$m? 
\end{enumerate}

HTRA science satisfies the fundamental E-ELT rationale of opening up new parameter space - the sub-second temporal domain. HTRA targets are examples of extreme physics - strong gravity (neutron stars \& black holes), highly relativistic plasmas ($\gamma > 10^9$) and strong magnetic fields (B $\approx 10^{15} G$ around magnetars).  Astronet's Panel A, developing A Science Vision for European Astronomy, identified six fundamental questions in the area of understanding extreme physics:

\begin{itemize}
\item How did the universe begin?
\item What is dark energy and dark matter?
\item Can we observe strong gravity in action?
\item How do supernovae and $\gamma$-ray bursts work?
\item How do black hole accretion, jets and outflows operate?
\item What do we learn from energetic radiation and particles?
\end{itemize}

Of these, the last four will benefit significantly from HTRA observations - indeed as in all these cases where there are variations on time scales from $\mu$seconds to seconds, it is not possible to understand their associated phenomena {\it without} HTRA. In Table 1 we summarise the existing observing timescales broken down by object and capability, now and in the ELT era. Other topics, possible with the E-ELT, include the remains of Population III stars, presumably now stellar mass black holes and searching for variability on the time scale of the last stable orbit combined with micro-lensing in globular clusters would be important tests of general relativity. Stochastic variability around young stellar objects (YSOs) could be indicative of periods analogous to the solar system's late bombardment period.

\begin{table}[h]
\begin{center}       
\begin{tabular}{|c|c|c|c|} 
\hline
\rule[-1ex]{0pt}{3.5ex}   &  & Time-Scale & Time Scale \\
\rule[-1ex]{0pt}{3.5ex}   &  & Now  & ELT era  \\
\hline
\rule[-1ex]{0pt}{3.5ex} Stellar flares  &  & Seconds/ & 10-100ms \\
\rule[-1ex]{0pt}{3.5ex} and pulsations  &  & minutes & \\
\hline
\rule[-1ex]{0pt}{3.5ex} Stellar  &  White Dwarfs &  1-1000 $\mu$s & 1-1000 $\mu$s\\
\rule[-1ex]{0pt}{3.5ex} Surface  &  Neutron Stars &  - & 0.1 $\mu$s \\
\rule[-1ex]{0pt}{3.5ex} Oscillations  &  &  &     \\
\hline
\rule[-1ex]{0pt}{3.5ex} Close Binary   & Tomography &  100ms++ & 10ms+     \\
\rule[-1ex]{0pt}{3.5ex} Systems  & Eclipse in/egress &  10ms+&  $<1$ms   \\
\rule[-1ex]{0pt}{3.5ex} accretion \& & Disk flickering  & 10ms & $<1$ms   \\
\rule[-1ex]{0pt}{3.5ex}  turbulence  & Correlations & 50ms &   $<1$ms   \\
\rule[-1ex]{0pt}{3.5ex}   &  (e.g. X \& optical) & & \\
\hline
\rule[-1ex]{0pt}{3.5ex} Pulsars  & Magnetospheric  & 1 $\mu$s- & ns  \\
\rule[-1ex]{0pt}{3.5ex}   & Thermal  & 100ms & ms \\
\hline
\rule[-1ex]{0pt}{3.5ex} AGN  &   & Minutes & Seconds \\
\hline
\end{tabular}
\end{center}
  \caption[science] 
   { \label{table:science} 
Science timescales showing current and future possibilities
}
\end{table} 

As well as reducing the timescale it will be possible to observe fainter objects, important in the context of neutron stars, as well as studies of temporally varying polarisation. HTRA is the only way to make a major step forward in our understanding of several important (astro)physical processes; these include the extreme gravity conditions around neutron stars and at the last stable orbit around stellar mass black holes. HTRA studies hence require astrophysical modeling and multi-frequency management. Very interesting applications lie in the regime of ultra-high time resolution, where quantum-physical phenomena, currently studied in laboratory physics, may be explored.

\subsection*{Pulsars and Neutron Stars}

Perhaps the most technically challenging HTRA targets are isolated neutron stars - either as radio pulsars, such as the Crab and Vela pulsars; radio quiet pulsars such as Geminga and PSR J0007+7303 discovered by the Fermi satellite; magnetars/anomalous X-ray pulsars (AXPs) such as 4U 0142+61, 1E 1048.1-5937 and recently SGR 0501+4516 \cite{dh10} as well as thermally emitting isolated neutron stars such as RX J185635-3754. Observationally the challenge comes from their intrinsic faintness (see Table 2) and their rapid variability. The Crab pulsar, for example, has been observed to show variations on nano-second time scales at radio wavelengths and $\mu$second timescales at optical wavelengths. To date, eight pulsars (4 normal, 1 radio-quiet and 3 magnetars) have been observed to pulsate optically, and a further sixteen have measured time integrated fluxes \cite{mi09}. In the E-ELT era we expect this number to rise significantly. 

\begin{figure}[htb]
\begin{center}
  \includegraphics[width=4in]{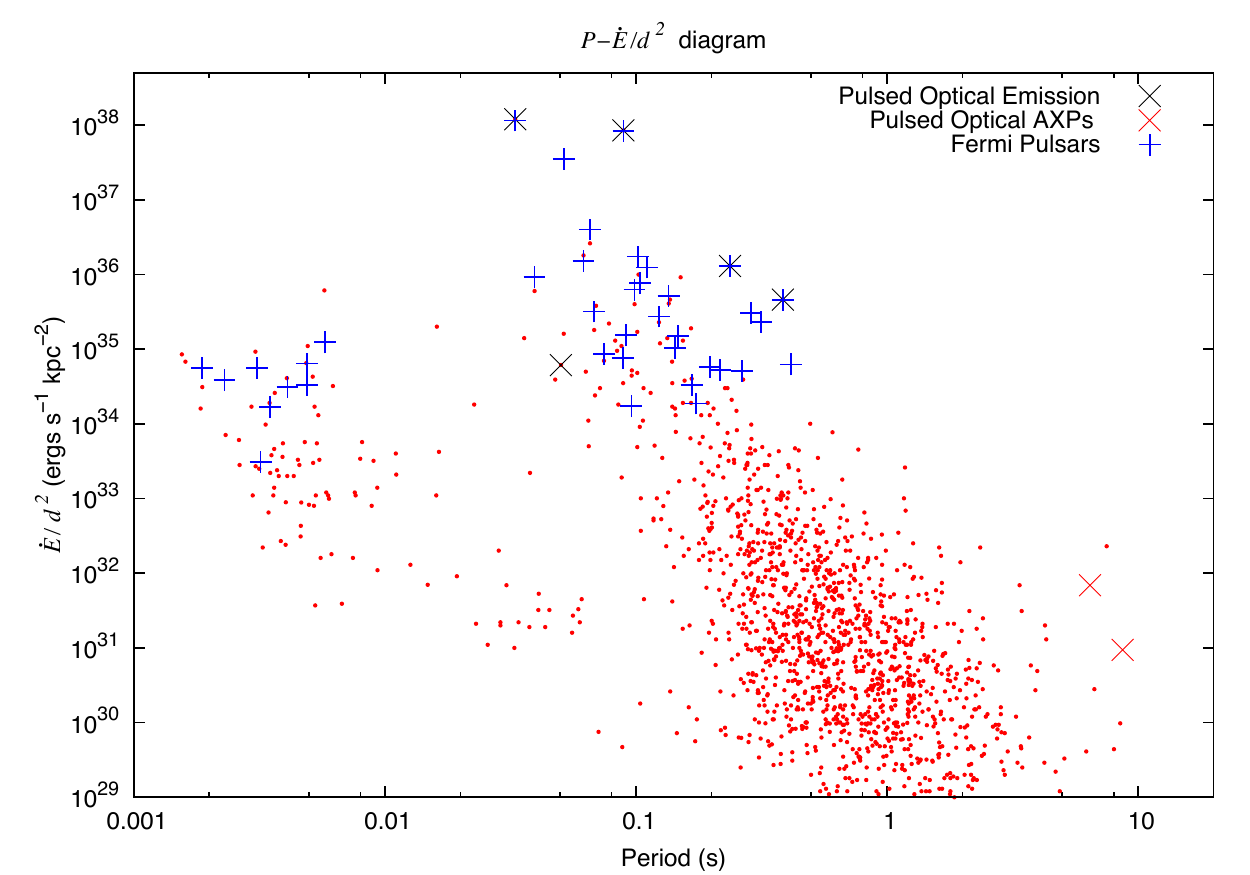} 
\caption{Period-Spin down energy diagram for radio (red points), Fermi(+) and optical pulsars($\times$) with the spin-down energy scaled for pulsar distance. The diagram indicates that Fermi pulsars have similar characteristics to pulsars with known optical counterparts. Only those Fermi pulsars with a known distance are displayed. We have shown with blue crosses likely optical pulsars, and *s show optical and $\gamma$-ray emitting pulsars.}
\label{default}
\end{center}
\end{figure}

Fig. 1 shows  pulsars with known optical pulsations; also indicated are the  radio pulsars and Fermi pulsars at known distances. The plot shows the pulsar period against pulsar spin down energy scaled for distance - a measure of the expected fluence.  

Although optical observations of pulsars are limited by their intrinsic faintness, it is in the optical that we have two distinct benefits with regard to other wavelengths - in the optical we can readily measure all Stokes parameters describing polarisation, and in the optical we are probably seeing a flux which scales linearly with local power density in the observer's line of sight. It is the latter which gives us important clues as to the nature of the emission zone for the high energy [h$\nu >$ 1eV] radiation from pulsars. The pulse shape and polarisation come from a mixture of the local physical conditions in the magnetosphere [magnetic field strength, plasma energy spectrum and pitch angle] and the relative geometry of the pulsar's rotation and magnetic field axes with the observer's line of sight. Changes in the optical pulse shape can be interpreted as changes in the location of the emission zone and hence the nature of the emission process. Changes in the optical polarisation give vital information about the local geometry of the high energy emission region. From the Fermi identifications we can reasonably expect over twenty more pulsars to be within the sensitivity of the E-ELT.

While timing is instrumental in securing the pulsar optical identification, the comparison of the optical light curves with the radio, X-ray, and $\gamma$--ray light curves is important to  determine the geometry of different emission regions in the neutron star magnetosphere and to test different emission models.  Moreover,  the study of the giant pulses in rotation-powered pulsars, so far observed only in the radio and in the optical bands \cite{sh03}, uniquely allows study of the relation between coherent and incoherent emission. For some pulsars which feature low radio fluxes, either because of their intrinsic radio faintness or because of their larger distance, optical observations are particularly important to study the evolution of the pulsar period through the measurement of the period first and second derivatives. This is the case, for instance, of the young pulsar PSR B0540-69 in the LMC (see Table 2) for which optical observations carried out over the last two decades allowed repeated measurements of the pulsar braking index (a crucial parameter which relates the pulsar frequency and its period derivative) to a very high level of accuracy, with the latest measurement obtained thanks to one of the latest generation HTRA instruments, IQUEYE \cite{na09}. Since several radio-silent pulsars are being discovered from X-ray and $\gamma$--ray observations, the measurement of the pulsar timing parameters in the optical domain becomes more and more important.

For magnetars, the comparison between pulsed fractions and relative phases of the optical and X-ray light curves  provides an important diagnostic to discriminate between emission models from the magnetosphere and from a surrounding debris disc created after the supernova explosion. In the case of the AXP 4U 0142+61  the optical pulsed fraction (29\% $\pm$ 8\%) was  a factor of $\approx$5  larger than the X-ray one, contrary to what is expected if the optical emission is due to  reprocessing of the magnetar X-ray radiation in a debris disc \cite{dh05}. At the same time, however,   no evidence was found for a significant optical-to-X-ray pulse lag, as would be, instead, expected in the reprocessing scenario\cite{dh05}. On the other hand, in the case of the AXP 1E 1048.1-5937 the optical pulsed fraction (21\% $\pm$ 7\%) was actually a factor of $\approx$0.7 lower than the X-ray one, as expected from X-ray reprocessing in a disc \cite{dh09}.  However, they also found  potentially significant evidence of an X-ray-to-optical pulse lag (0.06 $\pm$ 0.02), which is a counter-argument against the disc model. More timing observations of magnetars are important to verify the proposed emission models.

For the thermally emitting neutron stars, like RX J1856-3754,  the search for optical pulsations will be crucial to constrain the size of the emitting areas on the neutron star surface, which are found to be at different temperatures from the comparison of the optical and X-ray SEDs. This will allow a bias-free determination of the temperature  of the neutron star surface, and the study of heat transfer phenomena in the neutron star interior, together with the effect of the magnetic field.

\begin{table}[h]
\begin{center}       
\begin{tabular}{|c|c|c|c|c|} 
\hline
\rule[-1ex]{0pt}{3.5ex}  Object & Rotational & Magnitude &\multicolumn{2}{c|} {SNR 1 Hour Exp.} \\ \cline{4-5}
\rule[-1ex]{0pt}{3.5ex}   & Period (s) & (R/K Band) &  ~~~~10m~~~~  & 42m\\  
\hline
\rule[-1ex]{0pt}{3.5ex} Crab  & 0.033 & 16.8/13.8 & 7000 & 50000  \\
\hline
\rule[-1ex]{0pt}{3.5ex} Vela  & 0.089 & 23/20 & 200   & 800 \\
\hline
\rule[-1ex]{0pt}{3.5ex} PSR B540-69  & 0.050 & 24/21 & 100 & 320  \\
\hline
\rule[-1ex]{0pt}{3.5ex} PSR B0656+14  & 0.385 & 25/22 & 40 & 120  \\
\hline
\rule[-1ex]{0pt}{3.5ex} Geminga  & 0.257 & 26/23 & 15  & 45 \\
\hline
\rule[-1ex]{0pt}{3.5ex} 4U 0142+61  & 8.7   & 25/20 & 3   & 800 \\
\hline
\rule[-1ex]{0pt}{3.5ex} 1E 1048.1-5937  & 6.5  & 25/20 & 3  & 800 \\
\hline
\hline
\end{tabular}
\end{center}
\caption[detect] 
{ \label{table:detect} Summary of the currently known optical pulsars. The SNR column shows the expected signal to noise ratio (SNR)  from each object using 10m and 42m telescopes with a 7.5\% DQE instrument. For the 10m telescope we have assumed R band and natural 1 arcsec seeing and for the 42m, K band and 0.1 arcsec seeing.}
 
\end{table} 

In cases where an adequate spectral coverage is available through multi-band photometry evidence for multiple breaks has been found in the slope of the power-law (PL) magnetospheric emission continuum from the optical-UV to the IR and from the optical to the X-rays. Indeed, such spectral turnovers seem to be a common feature of rotation-powered pulsars \cite{mig10}, which probably suggests a complex particle distribution in the neutron star magnetosphere. This can be mapped through phase-resolved spectroscopy observations which would allow to unveil changes of the PL spectral index as a function of the spin period, thus unveiling differences in the emission process, corresponding to different projected regions of the neutron star magnetosphere close the magnetic poles. In many isolated neutron stars, thermal emission, presumably from a fraction of the neutron star surface larger and colder than the polar caps, is also observed in the optical-UV. The only way to accurately determine, coupled with the source distance, the neutron star surface temperature in the colder regions is by characterising the optical Rayleigh-Jeans (R-J) spectrum. Phase-resolved spectroscopy would also allow to measure the thermal spectrum evolution as a function of the neutron star rotation phase and to better locate emission regions at different temperatures. 

\subsection*{White Dwarfs}

As the most common end-product of stellar evolution, white dwarfs offer unique insights into stellar/binary formation and evolution. Diagnostics include cooling ages from effective temperatures, masses via pulsation seismology and accurate radii in eclipsing binaries. Exploding white dwarfs provide us with crucial Type Ia standard candles, and ultra-compact white dwarf binaries dominate the low-frequency gravitational waveband.

Dynamical time-scales near white dwarfs are of the order of seconds and thus moderately-high time resolution is needed to sample white dwarf phenomenology. Indeed, the early era of HTRA studies using photo-multipliers was instrumental for the discovery and development of the astrophysics of accretion in compact binaries thanks to pioneering studies of accreting white dwarfs.

High-time resolution photometry has been used to study white dwarf rotation periods (tens of seconds), pulsations modes (few hundred seconds), orbital modulations (mins to hours) and accretion-driven variability (seconds to months) around white dwarfs. Instruments such as ULTRACAM have permitted detailed studies of white dwarfs in compact binaries, resolving rapid eclipses and short orbital periods. These have produced, amongst others, accurate masses and radii for white dwarfs and their companions, tracked small period derivatives in the orbital evolution of white dwarf binaries and pulsation amplitudes as a function of wavelength. 

Cadence requirements are relatively modest compared to pulsars and neutron stars, but are still sufficiently short such that readout overheads and detector noise severely limit HTRA studies of white dwarfs with conventional instruments and telescopes. Even at exposures times of tens of seconds detector overheads often reduce the shutter-open duty cycle to below 50\% while readout-noise severely limits low S/N work despite large collecting areas. Given the scientific need to resolve timescales and limit exposure lengths accordingly, HTRA experiments are thus currently often forced to operate in low-efficiency configurations.

With large and relatively bright samples of targets in the optical, it is feasible to not just perform high time resolution multi-band photometry with resolution of seconds, but to push towards time-resolved high cadence spectroscopy in the era of ELTs and novel detector technologies such as Electron Multiplying CCDS (EMCCDs) and Scientific  CMOS (sCMOS). Readout noise limitations are an even larger concern in this case as the light is dispersed across many detector pixels. It is worth remembering that the regime of detector noise dominated data is not limited to high-time resolution work and thus improvements in this area will carry benefits well beyond HTRA.
\subsection*{Black Holes and Accretion Systems}

Black holes (BHs) in astrophysics are currently known to exist in binary systems (stellar mass BHs with masses up to several 10 M$_\odot$) and as supermassive BHs (SMBH, 10$^6$-10$^7$ M$_\odot$) in the nuclei of Active Galaxies. Intermediate mass BHs (hundreds of M$_\odot$) are expected to be responsible for ultra-luminous X-ray sources but have not yet been identified conclusively. The natural timescales around BHs, set e.g. by the orbital period of the innermost stable orbit, are typically in the millisecond range for stellar mass BHs (evidence of kHz QPOs) and in tens of minutes for SMBHs. Black hole X-ray binaries (BHXBs) are known to be powerful multi-wavelength emitters spending most of their life in quiescence but undergoing sporadic outbursts during which their X-ray luminosities can increase up to a factor of 10$^6$ compared to quiescent levels. In addition to the outburst light curves, the variability generated by the accretion flow into the BH and the formation of relativistic jets and outflows show highly variable emissions. The dominant emission from BH systems, especially from an accretion disk, is in the X-ray range and provides, through spectra and X-ray lines, fundamental diagnostics of the hot plasma in the inner disk.  Very strong and possibly beamed non-thermal radiation  (cyclotron, synchrotron, and inverse Compton) is also observed from relativistic jets. A close correlation of X-ray and optical emission has been found on various time-scales. During outbursts of the galactic BHXB GX 339-4, each lasting from weeks to several months, \cite{co09} found the striking correlation shown in Fig. 2 between the X-ray and V-band intensities.

\begin{figure}[htb]
\begin{center}
  \includegraphics[width=4in]{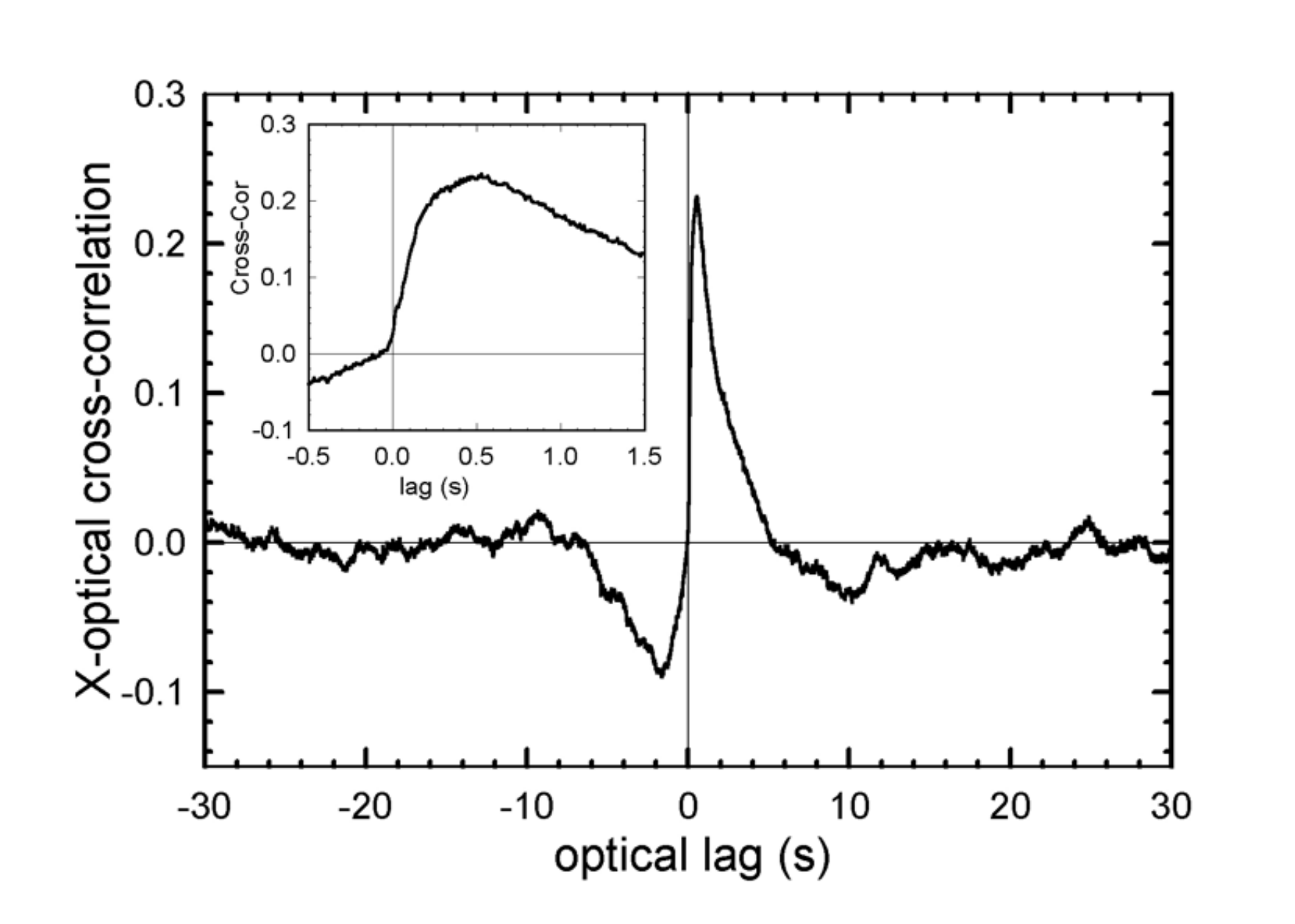} 
\caption{X-ray / optical cross-correlation function of fast variability in RX J1118+48 (=KV UMa) \cite{ka01}}
\label{default}
\end{center}
\end{figure}

Now focusing on high-speed simultaneous optical/X-ray photometry, three accreting BHs have shown phenomena that opened an exciting window to investigate the extreme physics near a black hole. Complex correlated variability in the optical and X-ray
emission was seen from RXTE J1118+480, while fast optical photometry of SWIFT J1753.5-0127 (Durant et al. 2008) and GX 339-4 \cite{ga08}  revealed further details. As examples, the cross-correlations between X-ray and optical variability for RXTE J1118+480 (Fig. 2) and GX 339-4 (Fig. 3) are shown. On second and sub-second timescales positive and negative responses of visible light to X-ray variability are detected. In the case of XTE J1118+480 \cite{ka01} explained the lag of about 500 ms between X-ray and optical response by a shock transit time in a sub-relativistic jet, while \cite{ma03,ma04} explained the behaviour through coupling of an optically emitting jet and an X-ray emitting corona to a common magnetic energy reservoir. An alternative explanation comes from the magnetically driven disc corona model \cite{me00}. In all models the correlations describe phenomena that occur less than 100 Schwarzschild radii from the BH.

\begin{figure}[htb]
\begin{center}
  \includegraphics[width=6in]{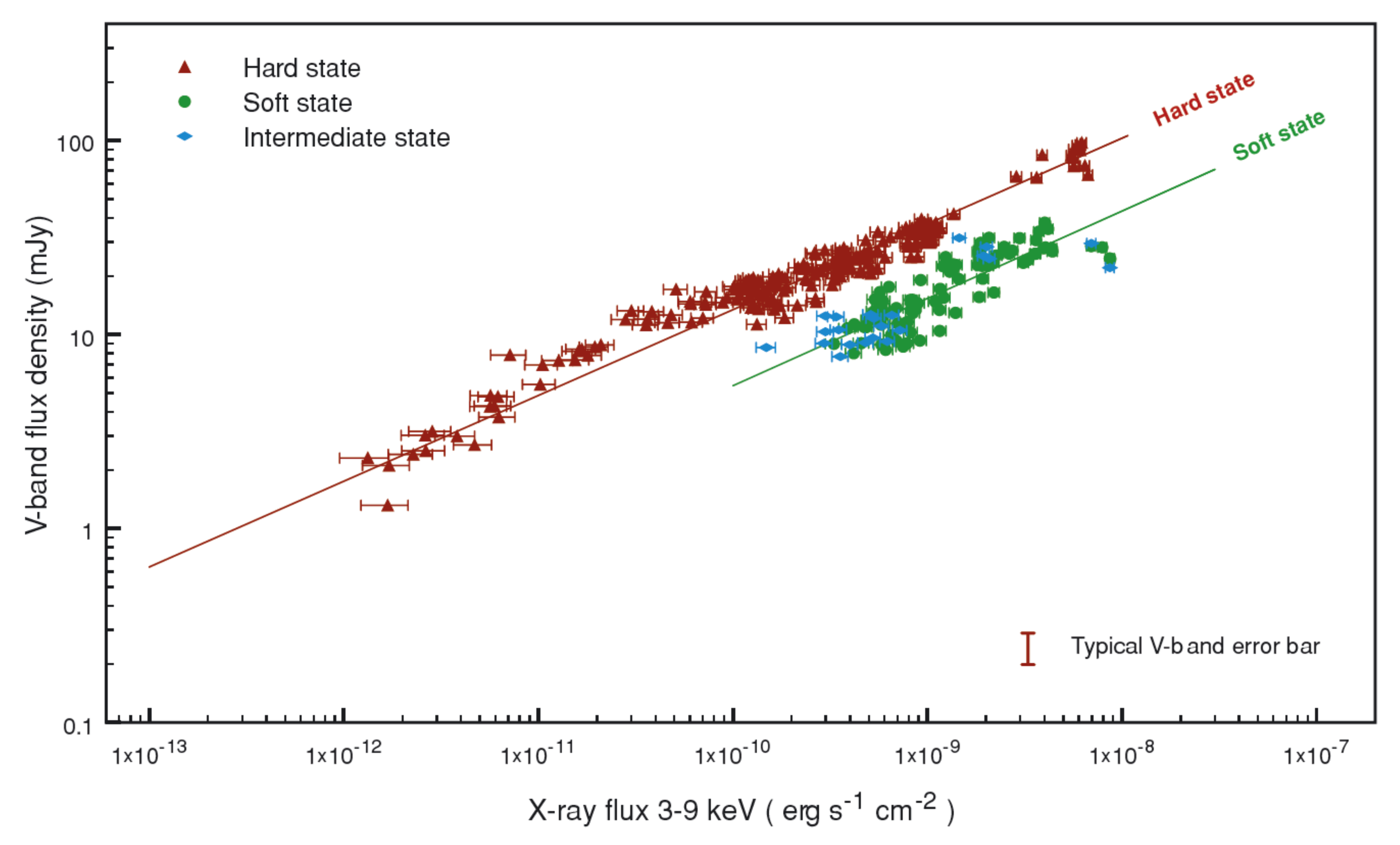} 
\caption{GX339-4: quasi-simultaneous optical V-band flux density versus 39 keV X-ray flux during four outbursts observed in 2002-2007. The colours indicate different X-ray states of emission\cite{co09} }
\label{default}
\end{center}
\end{figure}

High-time resolution NIR observations of GX 339-4 cast some new light on this issue, unambiguously unveiling variable synchrotron emission from the jet at time scales as short as tens of milliseconds (see Fig. 4). The measured time delay of 100 ms between the X-ray and the NIR jet emission allowed preliminary estimates of such physical parameters of the jet itself as the jet speed and the intensity of the internal magnetic field \cite{ca10}.  These data showed the exciting potential of NIR HTRA, now available only for relatively bright sources. Monitoring of the time delays between the NIR, optical and X-ray emission in the ELT era will allow to track the evolution of the jet parameters with the accretion rate, as well as their link with the accretion disk.

It is becoming increasingly clearer that multi-wavelength HTRA of accreting black holes is the best way to disentangle the physical origin of the complex broad-band spectral emission observed in these systems.
NIR and optical HTRA in the ELT era, possibly (but not necessarily) simultaneously with X-rays, will allow us to distinguish radiation from the different components of the accretion flow, each showing different variability properties.

\begin{figure}[htb]
\begin{center}
  \includegraphics[width=4in]{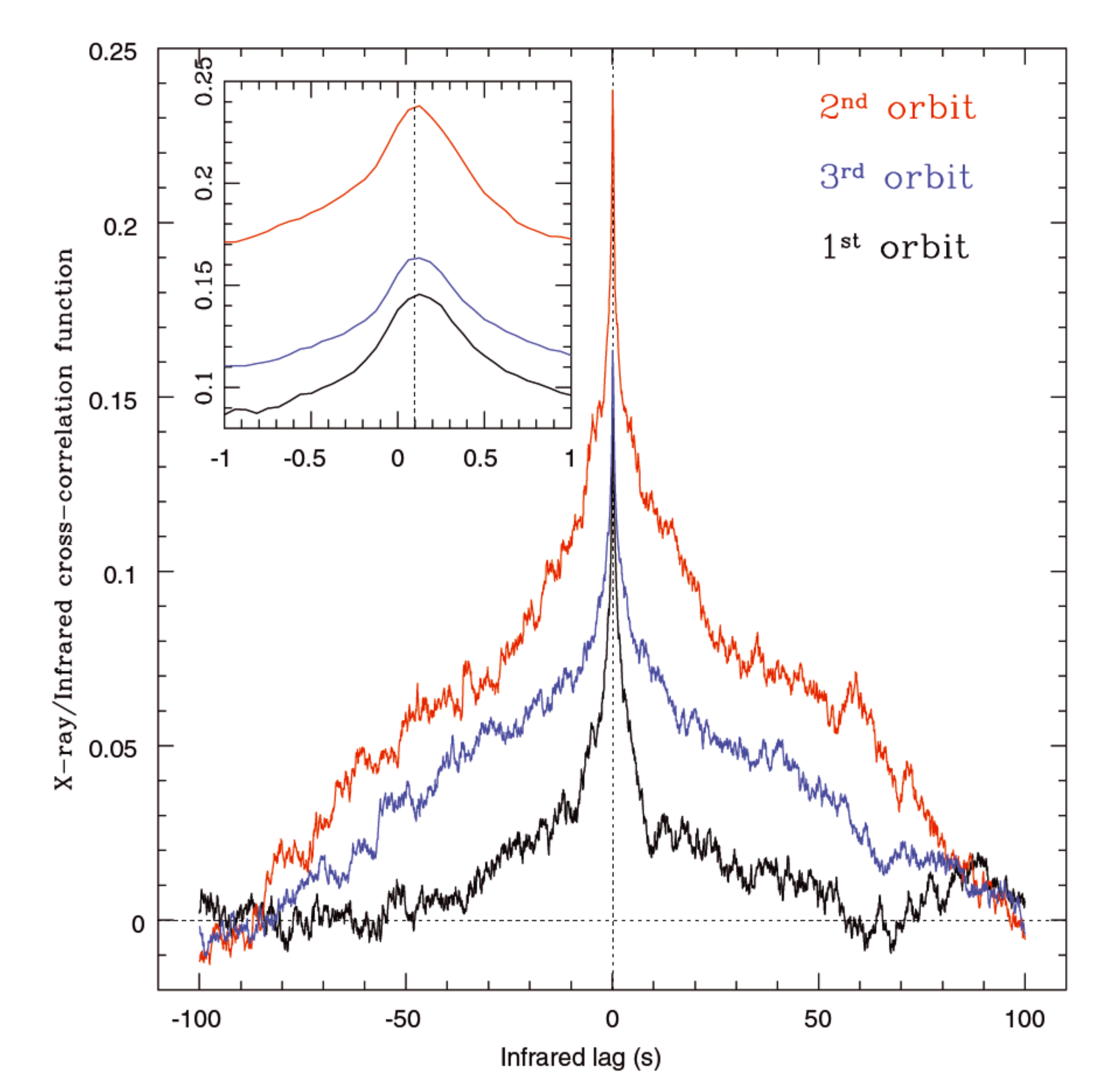} 
\caption{Cross-correlations of the X-ray and IR light curves of GX 339-4
(positive lags mean IR lags the X-rays). A strong, nearly symmetric correlation
is evident in all the three time intervals, corresponding to different
RXTE orbits. In the inset, we show a zoom of the peaks, showing the IR
delay of $\approx$100 ms with respect to the X-rays. The inset also shows a slight
asymmetry towards positive delays\cite{ca10}}. 
\label{default}
\end{center}
\end{figure}

In accreting BH systems we can thus distinguish optical radiation with different characteristics and possibly originating from different sites:
\begin{enumerate}

\item A general long term (scale of days to weeks) luminosity correlation with the X-ray emission during outbursts
\item Fast direct optical emission from regions in or above the accretion disk: flare-like cyclo-synchrotron emissions generated by stressed magnetic field configurations in a differentially rotating disk; self-absorbed synchrotron emission from a hot, magnetized corona. This emission could originate between 10 and 100 Schwarzschild radii from the BH as the range of variability time-scales for the X-ray emitting plasma is consistent with the dynamical times of an accretion flow in this region.
\item Reprocessed X-ray photons from the inner accretion disk scattering either on the outer disk material or on the photosphere of the secondary, donor star (light-echoes).  The first examples of these light echoes have been already observed in a few BH systems \cite{ob02,hy09}, although observations are now limited by the need for bright targets. HTRA with the ELTs will allow the extension of these experiments to much fainter sources, providing clear determinations of the size of many systems. This will yield the first population statistics of accreting black holes in binary systems, allowing a major step forward in the study of binary evolution.

\item Light generated by energetic electrons in a magnetized jet or outflow through cyclo-synchrotron emissions. Jets have so far only been imaged in so called micro-quasars by radio interferometry. Well-known examples of micro-quasars containing BHs are GRS 1915+105 and the very bright X-ray source Cyg X-1. Optical photometry and polarimetry could provide essential insights into the formation of jets in BH binary systems.
\end{enumerate}

Future HTRA facilities, hopefully extending the optical coverage further into the NIR/IR range and making use of the huge photon collection area of E-ELT, can provide the reach to study black hole binary systems at much larger distances, in shrouded environments, and possibly at faint emission states. The synergies with future X-ray (NuStar, EXIST, IXO) and radio facilities (SKA) will further advance our understanding of these extreme astrophysical systems.

\subsection*{Stellar Phenomena}
Normal stars undergo a number of changes on time scales of a few seconds. These vary from stellar pulsations, where the whole star changes in size and brightness on regular or quasi-periodic timescales, to stochastic explosive phenomena indicative of intense magnetic activity in the stellar atmosphere. Radial and non-radial pulsations provide unique opportunities to examine the internal structure of stars using the techniques of astro-seismology. Pulsations may also provide the trigger for large mass loss seen in many luminous stars. Moreover, pulsations are associated with transient atmospheric shocks that affect the structure of the stellar atmosphere. With large telescopes and new instruments capable of high time resolution spectroscopy and multi-colour photometry, we aim to extend these techniques from simple frequency analyses of a few bright stars to in-depth analyses of pulsation modes and amplitudes in a variety of stars hitherto inaccessible. Primary objectives will be the sub-dwarf B stars and $\delta$ Scuti stars. 

The study of flares in all stars, including the Sun, is hampered by their unpredictability and their short duration. An entire explosive event may last only a few hundreds of seconds, with most of the crucial physics occurring within a few seconds of outburst. High time-resolution observations are required to understand the physics of explosive events on stellar surfaces, and to establish how they scale from the Solar network to the mega-flares seen on very cool stars.

With the advent of optical interferometers and very large telescopes, stellar astrophysics is facing a major transformation in that many stars are now becoming spatially resolved surface objects (rather than the point sources of the past). However, to spatially resolve smaller stars will still not be possible with imaging methods. Nevertheless, HTRA may permit indirect studies of their surface structure. For example, hydrodynamic models of white-dwarf surfaces show a pattern of stellar surface features (corresponding to solar granulation) with high velocities (tens of km/s), and short lifetimes (a fraction of a second). Although the number of such features across any stellar surface is large, the number is finite, causing slight photometric flickering in integrated starlight, potentially detectable with HTRA instruments at large telescopes.

\subsection*{Occultations}

Precise timing of stellar occultations of the icy moons of Saturn: In 1 ms, Enceladus moves by 52.2 m along its orbit, and timing its occultation to that precision gives an astrometric accuracy approximately 10 times better than with Cassini. These considerations can be extended to other moons of Saturn. This precision, unattainable by any other means, including spacecraft, will allow a better knowledge of the dynamics of the Saturnian system. 

Lunar (LO) and Kuiper Belt  Object (KBO) Occultations: The method of observing lunar occultations of background stars to derive their angular diameter, as well as binarity and circumstellar matter, provides angular resolution at the milliarcsecond level independently of the diameter of the telescope used. This surpasses the diffraction limit of even the largest single telescopes of the next generation and rivals the resolution of long-baseline interferometry. In particular, the LO technique at a very large telescope is a powerful and efficient method for achieving angular resolution, sensitivity, and dynamic range that are among the best possible today with any technique \cite{ri08}. The selection of targets is naturally limited and LOs are fixed-time events; however, each observation requires only a couple of minutes, including overheads. As such, LOs are ideally suited to filling small gaps during the idle time between other observations. The same can be said of stellar occultations by KBOs. In this case, the method will provide information on KBO characteristics, such as diameter or atmosphere, impossible or very difficult to measure by other means. Another HTRA requirement comes from studying the diffraction pattern associated with KBO occultations \cite{ro00}.

\subsection*{Transits}

Transiting planets play a key role to the understanding of the formation and evolution of planetary systems. When photometric data are coupled with radial velocity data, it is possible to derive mass radius and density of the planet, and hence its internal structure can be understood. To date, space and ground surveys of transiting planets (Corot, HAT, Kepler, OGLE, SuperWASP, TrES, XO) have discovered about 60 planets (http://exoplanet.eu), and the number is constantly increasing. The vast majority of these planets are hot Jupiters, with orbital periods of few days. Few Neptune-mass transiting planets (like Corot 7b, GJ 436b) are also known. Detection of super-Earth planets could be performed measuring transit time variations (TTV) or transit duration variation (TDV). The measurement of the mid-transit times with high temporal resolution, can be used to detect the presence of a third body perturbing the orbit of the known transiting planet. This technique is more sensitive to resonant orbits of the third body, as the induced variations are the largest. For the detection of these planets, mid-transit times need to be measured with an accuracy better than 10 seconds \cite{sa99}. Current knowledge of transit mid-time for the three planets proposed for observations is around 15 s (Corot-212 s, \cite{al08}, WASP-2 20 s \cite{ca07}, HD189733b 24 s\cite{mi08}. Analysis of HD 189733b transits obtained with MOST satellite \cite{mi08}  have excluded only the presence of very massive perturbers and planets of masses larger than 1-4 solar masses. A search for Trojan planets in known transiting systems was performed by \cite{ma09}  combining photometric and radial velocity data. The authors obtained upper limits on mid transit-time variations of the order of several minutes for each planet analyzed. To exploit this possibility, several instrumental features are needed: high quantum efficiency, high temporal resolution, high stability and high accuracy clock running for hours, and acquisition devices capable of sustaining high photon rates. 

\subsection*{Asteroids}

While stars and planets are
more or less spherical, asteroids are non-spherical. After space
missions, the best known way to study their shape and surface features
is through planetary radar, which is applicable only to a small subset of
objects (near and large enough). For all other objects, the main tool
is time resolved photometry. Especially interesting are the rare so-called
'tumbling asteroids', whose non-principal-axis rotation reflects some
perturbation that happened relatively recently \cite{pr05}. Phase resolved photometry,
and maybe even phase resolved spectroscopy and polarimetry \cite{ta08,mu02} can be very
valuable in the study of the dynamical interactions in the Solar System,
as well as the material properties of asteroids. A related phenomenon,
which requires similar observations, and contributes similar science,
is the phenomenon of eclipsing binary asteroids \cite{no05}.

\subsection*{Polarisation}

Many HTRA targets are non-thermal emitters with emission dominated by the interaction of particles with strong magnetic fields. As such polarisation plays a key role in determining the nature of the source function of the objects. In this regard the design of HTRA instruments should take into account polarisation when considering the focal station for the instrument. 

\subsection*{Polarimetric observations of pulsars}
High time resolution polarisation observations of pulsars play a key role in understanding the emission process that takes place in their magnetospheres. Neutron stars, pulsars and magnetars are by far the best targets for time resolved polarimetry on the time scales of few microseconds. Among all celestial sources pulsars and magnetars have the highest magnetic fields. This implies a high degree of polarisation. In addition, the fast rotation of these objects causes drastic changes of the polarisation degree (P.D.) and position angle (P.A.) on the time scales of few hundreds of microseconds \cite{sl09}, Fig 5. Precise measurements of P.D. and P.A. as a function of pulse phase in a wide range of energy domains would provide deeper insight into the details of the pulsar emission mechanisms. However, since fast X- and $\gamma$--ray polarimetry is unavailable so far, we are restricted to the optical domain. Yet, despite the increasing number (24) of  pulsars with optical counterparts, an attempt to measure their intrinsic and/or their nebulae polarisation characteristics was made only for five of them.

In the radio domain, high time resolution polarimetry measurements on time scales of few nanoseconds have been  performed \cite{ha03}. The situation is different in the optical domain.So far, only for the Crab pulsar has the optical polarisation characteristic been resolved down to 10 microseconds. These observations show a strong connection between coherent radio and non-coherent optical emission, i.e. there is a sudden change in position angle not only for
the phase of optical maximum, but also for the phase where the maximum of radio emission is observed. This observational evidence is in a strong agreement with other simultaneous radio and optical observations showing the enhanced optical emission while the giant radio pulses occur \cite{sh03}

\begin{figure}[htb]
\begin{center}
  \includegraphics[width=4in]{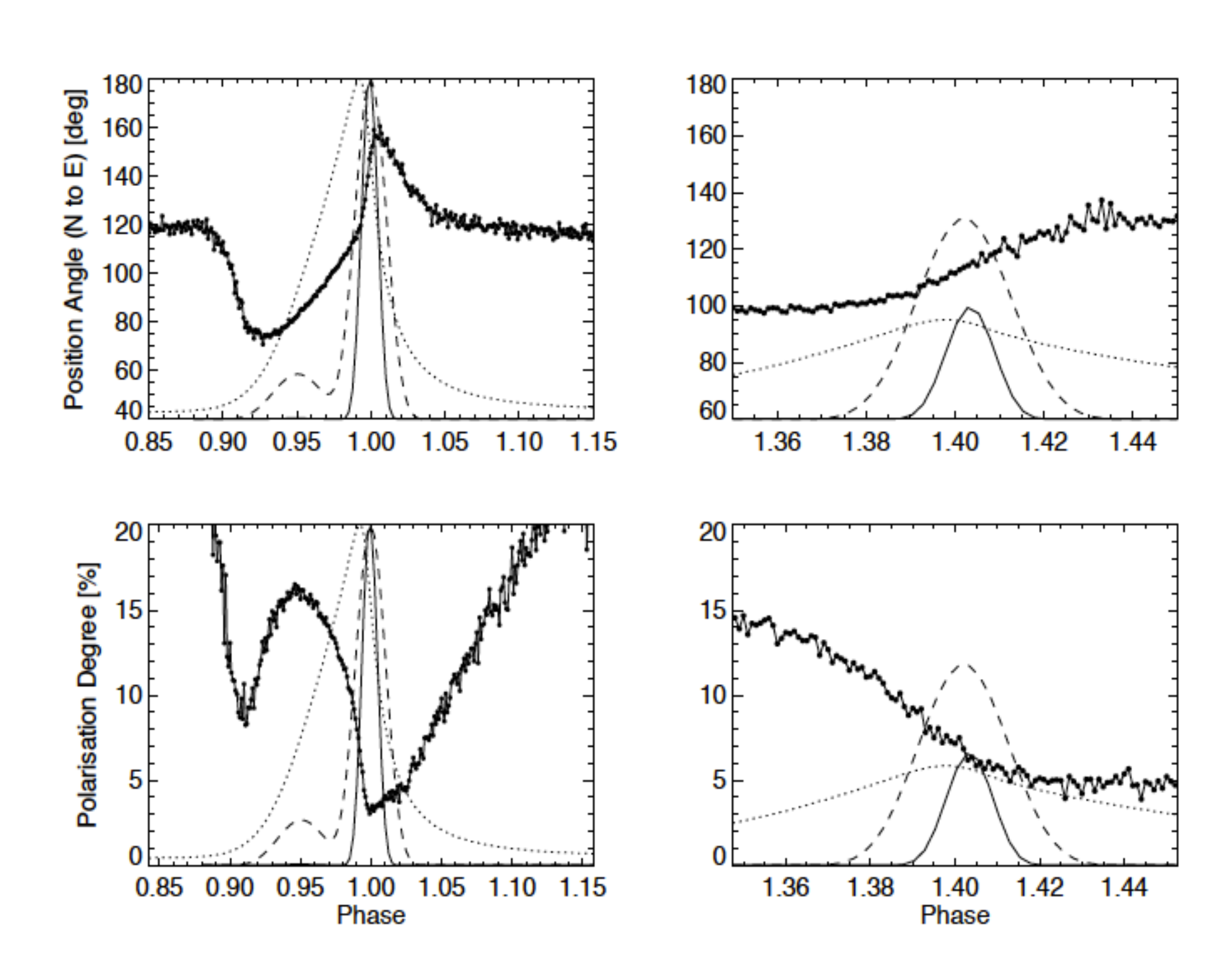} 
\caption{From \cite{sl09}  their Figure 8 - Optical polarisation characteristics of the Crab pulsar (P.A. - top row, P.D. - bottom row) compared with the pulsar radio profiles. Radio profiles obtained at radio frequencies of 1400 MHz and 610 MHz are shown as a solid line and dashed line, respectively. Left column shows zoom around the main pulse phase, whereas right column around the inter pulse phase. Points indicate the optical polarisation measurements, while the dotted line shows the optical intensity profile. This figure shows the unexpected result that the peak of the optical polarisation is coincident with the radio precursor and not the optical pulse.}
\label{default}
\end{center}
\end{figure}

The models of pulsar magnetospheric activity are based on various assumptions and boundary conditions. These lead to different models that describe the macro (spatial extent of accelerators and emitting regions), as well as the micro scale (specific radiative processes). The macro scale models include polar, slot, caustic and outer gaps as well as striped winds. The latter includes e.g. curvature radiation, synchrotron radiation and inverse Compton scattering. In consequence, the models differ significantly in the resulting `observed' radiation properties: light curves, energy spectra, and - last but not least - polarisation. Some models, are able to reproduce, albeit very roughly, some of the observed optical properties.

\cite{ta07} attempted to reproduce simultaneously all known high energy emission properties of the Crab pulsar, including its optical polarisation characteristics, within the framework of the modified outer gap model. The model is restricted to synchrotron emission due to secondary and tertiary electron-positron pairs which are expected in different spatial locations of the 3-D gap. Internal polarisation characteristics were calculated with particular care. Yet, the calculated polarisation properties of optical emission hardly reproduced the observed properties. However, like in the case of the two-pole caustic model \cite{dy04,dy04b} the polarisation characteristics obtained by \cite{ta07} become more consistent with the Crab pulsar optical data only after the DC subtraction. An important outcome of \cite{ta07} is the energy dependence of the polarisation features, covering the energy range between 1eV and 10keV. Together with the spectral and photometric fingerprints, polarimetric characteristics will provide additional constraints to the
existing and future models. It is in the optical and NIR regime, where all the characteristics of the radiation will be accessible, that we expect that we can separate these models  and hence understand the pulsar emission process.

\subsection*{Polarimetric observations of magnetars}

The magnetars should be definitely considered as distinguished candidates for high time resolution polarimetry. They are believed to be neutron stars with extremely high dipolar magnetic fields, and their activity powered by magnetism \cite{du92, pa92}. Up to now, at least two magnetars were observed in the radio range and recently there appeared one possible optical magnetar candidate \cite{st08}. There arises a question whether or not the optical and radio activity are also powered by magnetism. It is well known that in the magnetosphere of a classical pulsar such kind of radiation is generated by the relativistic particles accelerated in inner and/or outer gap and the necessary energy comes from the neutron star spindown energy loss. The crucial point would be the optical detection of magnetars during their X-ray activity. Inasmuch as their rotation period is in a range of few seconds, high time resolution is required for these observations. If we are able to measure polarisation and position angle direction in both radio and optical ranges then we
will get some constraints on the radiation mechanisms. 

We could expect that optical radiation is connected with synchrotron process, while for the radio emission the coherency is necessary. On the other hand both of them must occur in magnetospheric plasma. However, while they propagate in the magnetosphere they should interact with plasma, therefore their polarisation characteristics should be defined by ambient plasma and magnetic field configuration. If we assume that the magnetosphere of magnetars consists of the relativistic magnetized electron-positron plasma, as it is expected for classical pulsars, then the electromagnetic waves should naturally split into normal modes of plasma, i.e. ordinary and extraordinary modes. Comparison of measured phased-resolved position angles of radio and optical emission will provide us with extremely important information about radiative processes in magnetars. If the position angles of radio and optical emission are parallel to each other then both of them should be extraordinary modes, while if no correlation is observed then one can conclude that the waves are originated in different regions of the magnetosphere. Thus the high time resolution optical photo-polarimetry of magnetars is important not only for magnetar physics but also generally for classical pulsars.

\subsection*{Polarisation observations of close binary systems}

In the last few years it has been discovered \cite{sh08,ru08} that the polarisation signature of synchrotron emission from the jets of accreting stellar-mass BHs and neutron stars can be detected in the IR, and is variable on timescales of minutes. The level of polarisation implies a predominantly tangled, only slightly ordered magnetic field near the base of the jet in these systems. Recent results\cite{ru10}show that the polarisation flares from the jet in Sco X-1 can also be detected at optical wavelengths, on shorter timescales -- seconds (see Fig. 6). The magnetic field at the location close to the jet base is therefore turbulent, rapidly changing on timescales at least as short as 10 seconds. Sco X-1 is the brightest low-mass X-ray binary; only fast-polarimeters on extremely large telescopes can uncover the true rapidity of jets from these systems in general, in both flux and polarisation, and directly uncover the answers to the Astronet's Science Vision question `How do black hole accretion, jets and outflows operate?'

\begin{figure}[htbp]
\begin{center}
  \includegraphics[width=6in]{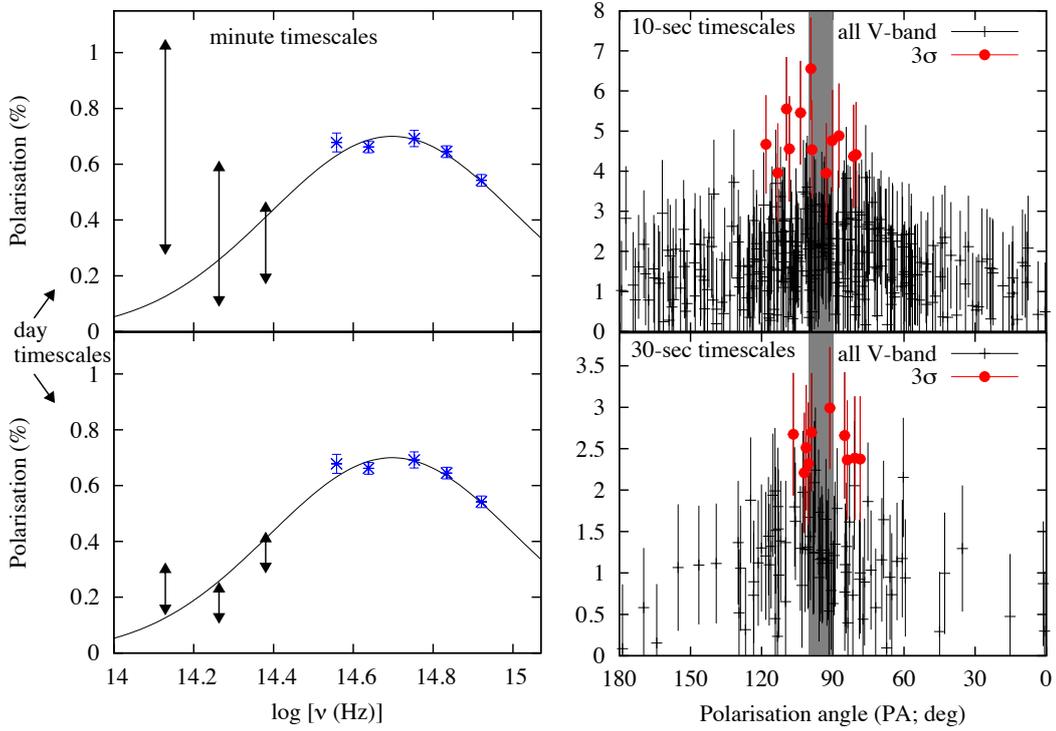} 
\caption{Linear polarisation of Sco X-1 on various timescales. left panels: polarisation versus frequency for optical results (blue crosses)\cite{sh04}   with model fit for an origin of interstellar dust (black curve), and IR results from \cite{ru08} taken with UKIRT. The black arrows indicate the range of values measured on each of the two dates (upper and lower panels). In the upper left panel significant intrinsic polarisation was detected which was variable on timescales of minutes and is consistent with a synchrotron jet origin. Right panels: polarisation versus polarisation angle (PA) for a shorter timescale variability study of Sco X-1 at optical wavelengths \cite{ru10} taken with HIPPO on the SAAO 1.9-m telescope. On 10 and 30-second timescales, short, significant ($>$ 3$\sigma$) flares of polarisation are detected, with the same PA as was seen from the jet in the IR (grey region).}
\label{default}
\end{center}
\end{figure}

\subsection*{Intensity interferometry and Photon Correlation Spectroscopy}The fastest variability of any light source is that caused by the quantum nature of radiation, on timescales equal to the coherence time of light. One astrophysical application is intensity interferometry, pioneered by R. Hanbury Brown and collaborators, measuring the second-order spatial coherence of light by cross correlating the very fast (nanoseconds) intensity fluctuations simultaneously measured in two spatially separated telescopes.  By combining such measurements over baselines of different lengths, an image of the source may be reconstructed, somewhat analogous to the ordinary phase (amplitude) interferometry.  One great advantage of intensity interferometry is that it is (virtually) insensitive to atmospheric turbulence or optical imperfections in the telescope, and therefore can be applied over long baselines, at short optical wavelengths and when observing through large air masses.  A requirement, however, is a large photon flux, and thus it requires bright sources and large telescopes. For the E-ELT, this means that diffraction-limited imaging in the blue or violet should be feasible for at least brighter stars, with a rather simple multichannel ultra-high-speed photometer that in software correlates signals between different segments of the main mirror. Such an instrument (which does not require good seeing, nor adaptive optics, nor a filled telescope aperture) was evaluated within the QuantEYE instrument study made already for the OWL telescope, and the concept has since been verified by the Iqueye visiting instrument used at ESO in Chile. Other physical observables can be extracted from nanosecond photon statistics\cite{leb06,dr08}.  A somewhat related method to the above can be used for measuring the second-order temporal coherence of light, reaching a spectral resolution orders of magnitude beyond what is feasible with hardware spectrometers.  Such intensity-correlation spectroscopy might be used to search for extremely narrow optical emission lines, e.g., natural lasers in sources such as Eta Carinae.

\subsection*{Current HTRA Instrumentation}

Table 3 gives a summary of the existing HTRA instruments.

\begin{table}[h]
\begin{center}       
\begin{tabular}{|c|c|c|c|c|c|} 
\hline
\rule[-1ex]{0pt}{3.5ex}  Detector & Time & Quantum & E/$\Delta$E & No. of & Instrument \\
\rule[-1ex]{0pt}{3.5ex}   & Resolution & Efficiency &   & Pixels &  \\
\hline
\rule[-1ex]{0pt}{3.5ex} CCD  & 5ms+ & 90\% + & -  & $>>10^6$ & UltraCam\cite{dh07}\\
\hline
\rule[-1ex]{0pt}{3.5ex} EMCCD  & 1ms+ & 15\% + & -  & $10^6$ & UltraSpec\cite{dh07} \\
\rule[-1ex]{0pt}{3.5ex} EMCCD  & 1ms+ & 15\% + & -  & $10^6$ & GASP\cite{co10}\\
\hline
\rule[-1ex]{0pt}{3.5ex} pn CCD  & 0.01 ms+ & 90\% + & -  & $10^6$ &  \cite{ha08}\\
\hline
\rule[-1ex]{0pt}{3.5ex} Active Pixel  & a few $\mu$s & 80\% + & -  & $10^5$ & \cite{fi06} \\
\rule[-1ex]{0pt}{3.5ex} Detectors  &  &  &  &  &  \\
\hline
\rule[-1ex]{0pt}{3.5ex} SPADs  & ns+ & 80\% + & -  & a few & Optima\cite{ka03}\\
\rule[-1ex]{0pt}{3.5ex}  & ns+  & 15\% &   &  one$^a$ & GASP\cite{co10} \\
\rule[-1ex]{0pt}{3.5ex}  &  100ps & 50\%+ &  -  & a few  & Iqueye\cite{ba07}\\

\hline
\rule[-1ex]{0pt}{3.5ex} STJ  & ns+ & 90\% +& 5  & 10s & SCAM\cite{ve06}\\
\rule[-1ex]{0pt}{3.5ex} TES  & ns+ & 90\% + & 20+  & 10s & \cite{ba06} \\
\rule[-1ex]{0pt}{3.5ex} MKID  & ns+ & 90\% + & 500+  & 10s & \cite{ma10} \\
\hline
\rule[-1ex]{0pt}{3.5ex} Photo-  & ns+ & $<$30\%  & -  & $1-10^6$ & Many \\
\rule[-1ex]{0pt}{3.5ex} Cathodes  & 1ms & 40\%  & -  & $10^6$ & wavefront \\
\rule[-1ex]{0pt}{3.5ex}   &  &  &   &  & sensor \\
\hline
\end{tabular}
\end{center}
\caption[detect] 
   { \label{table:detect} Current detector types showing the effective number of pixels, time resolution, DQE and energy resolution. \\
   Note: Galway Astronomical Stokes Polarimeter (GASP) has 4 channels for determining polarisation, the QE quoted is per channel. These instruments are mainly private rather than common user instruments. The exception is UltraCam and UltraSpec, which are available as a semi-common user instrument. 
   }
\end{table}

\subsection*{Detector Possibilities to 2.5 microns}

ELTs present the possibility of opening up HTRA science, however there are a number of significant problems extending HTRA to 2.4 microns. Most existing detectors are silicon based which stops being effective at about 1 micron. A number of possibilities are emerging in the NIR  from 1-2.5 microns including superconducting detectors such as Transition Edge Sensors(TES) devices \cite{ba06}, Superconducting Tunnel Junctions (STJs) \cite{ve06} and microwave kinetic induction devices (MKIDs) \cite{ma10} which are sensitive down up to at least 2 microns; InP single photon Avalanche Photo-Diodes (SPADs) sensitive between 1 and 1.6 microns \cite{it08}  and HgCdTe hybrid devices \cite{fi06} The history of superconducting devices has shown that they are difficult to operate in an observatory requiring liquid helium cryostats and have had significant stability problems. This is alleviated to some extent with MKIDs, in which the detector needs liquid helium, but the electronics do not. In the NIR HgCdTe Avalanche Photo-Diode (APD) arrays  offer possibly the best option and moderate sized arrays will become available over the next 2-3 years \cite{ha09}. For exposure time greater than 50ms conventional CCD systems are currently available for NIR observations with noise levels down to 4e$^-$/pixel, below the sky noise limit. Table 4 shows the expected sky counts for different seeing conditions with a 42m telescope. The numbers have been extrapolated from the ESO E-ELT exposure time calculator. 

\begin{table}[h]
\begin{center}       
\begin{tabular}{|c|c|c|c|c|c|} 
\hline
\rule[-1ex]{0pt}{3.5ex}  Pass & Wavelength & Sky Brightness &  \multicolumn{3}{c|} {Counts/millisecond}   \\
\rule[-1ex]{0pt}{3.5ex}   & &  &  \multicolumn{3}{c|} {within aperture}   \\ \cline{4-6}
\rule[-1ex]{0pt}{3.5ex}  Band & (microns) & (Jy arcsec$^{-2}$) &   1"  & 0".2 & 0".02\\
\hline
\rule[-1ex]{0pt}{3.5ex}  J & 1.25 & 6230 &   6.23  & 0.25 & 0.003 \\
\hline
\rule[-1ex]{0pt}{3.5ex}  H &  1.65 & 25000 &   25.0  & 1 & 0.01 \\
\hline
\rule[-1ex]{0pt}{3.5ex}  K & 2 & 34250 &   34.2  & 1.3 & 0.01 \\
\hline
\rule[-1ex]{0pt}{3.5ex}  L & 3.45 & 9.7$\times 10^9$ &   9716  & 400 & 4.0 \\
\hline
\end{tabular}
\end{center}
\caption[detect] 
{ \label{table:counts} Sky counts for given sky apertures of 1", 0".1 and 0".01. This shows that for seeing better than $\approx$ 0".2 that very low noise
 detectors with read noise of $< 1e^-$ / frame or integration time will be needed.  }
 
\end{table}

\vskip 0.2cm
\subsection*{Possible HTRA Instrument Configurations for ELTs}

HTRA instruments are generally small with simple optics. In reality it is only the detector and associated software which is important. As such, all instruments are in principle HTRA instruments given the correct detector. However the fast-timing aspects are often introduced into instrument design as an after-thought, so there is either limited efficiency due to readout problems or insufficient hardware and software for high-speed data recording. A possibility, for the E-ELT, is to include an HTRA instrument within an existing instrument. An example of this is the proposal for a simple HTRA instrument within the Micado cryogenic enclosure. We have also looked at and are in discussions with the various fibre fed instruments with a view to have a few fibres linked to an HTRA instrument that could then be located away from the main instrument - an example of this would be Optimos. This also has the advantage of both an optical and NIR capability.  Fundamentally the HTRA capabilities of the E-ELT are a detector and software issue; we would ask that HTRA is considered at the design stage of instruments so that absolute timing, read-out, storage and preprocessing can be properly addressed.

\subsection*{Authors and Acknowledgements}

The document was prepared by members of the Opticon HTRA network - A. Shearer, G. Kanbach, A. S{\l}owikowska, C. Barbieri, T. Marsh, V. Dhillon, R. Mignani, D. Dravins, C. Gouiff\'es, C. MacKay and S. Collins. We also acknowledge contributions from participants at the HTRA IV workshop held in Crete from 5th-7th May 2010 (see www.htra.ie) specifically D. Russell, P. Casella, S. Zucker, A. Rau, A. Zajczyk, D. Steeghs  \& V. Trimble.

\end{document}